# Reducing Average and Peak Temperatures of VLSI CMOS Digital Circuits by Means of Heuristic Scheduling Algorithm

Wladyslaw Szczesniak

Faculty of Electronics, Telecommunications and Informatics
Gdansk University of Technology, 80-952 Gdansk, Poland

*Abstract-*This paper presents a BPD (Balanced Power Dissipation) heuristic scheduling algorithm applied to VLSI CMOS digital circuits/systems in order to reduce the global computational demand and provide balanced power dissipation of computational units of the designed digital VLSI CMOS system during the task assignment stage. It results in reduction of the average and peak temperatures of VLSI CMOS digital circuits. The elaborated algorithm is based on balanced power dissipation of local computational (processing) units and does not deteriorate the throughput of the whole VLSI CMOS digital system.

## I. INTRODUCTION

This paper concerns application of heuristic scheduling algorithm to balance the load of tasks onto computation units ($cu_i$) uniformly with reduction of the total cost of the digital VLSI CMOS system ($co$) but without deteriorating its global computational efficiency measured e.g. by its throughput ($Th$). In this paper, the universal measure, named the cost ($co$) represents the consumption of power supply of VLSI CMOS digital system.

In the literature we can find a variety of methods concerning computational task assignment to different computation units ($cu_i$) [1], [2], [4], [5], [11]. It is especially important for reducing the supply power demand [6], [7], [8], [9]. In the paper the design objective function taken into account is the cost of the system ($co$). This measure has the straightforward influence on average and peak temperature of IC.

The presented BPD (Balanced Power Dissipation) algorithm can be applied to reduce global computational demand and provides balanced power dissipation of the digital VLSI CMOS system at the task assignment stage.

## II. PROBLEM FORMULATION

The task to be computed is described by the tasks' graph $G_T(V,E)$ as presented in Fig.1. An individual computational task ($ct$) in graph $G_T$ is represented by vertex $v_i$ in the set of vertices (e.g. in Fig. 1 ☼, ▲, ●). Each $ct$ has to be assigned to one of the computational units of a given resource type

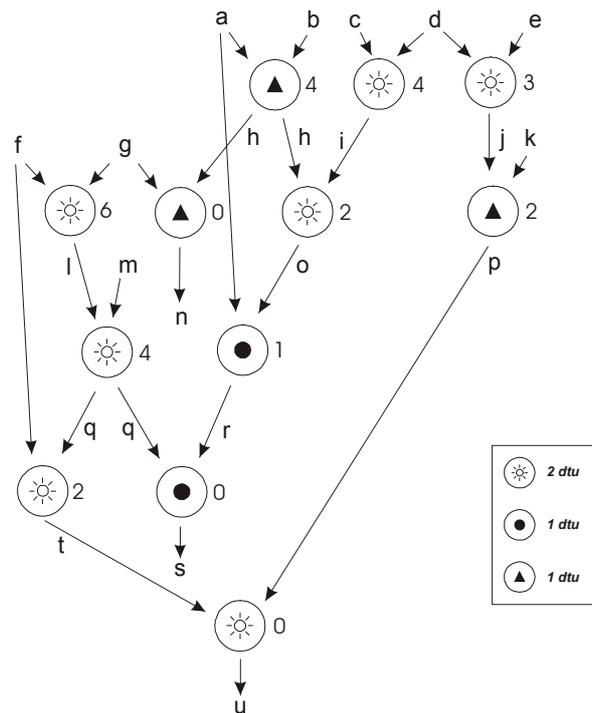

Fig. 1. An example of a computational task represented by the tasks graph GT (where a, b, c, d, e, f, g, k, m represent the sets of input data) annotated with *p*-labels (the number of discrete time units dtu for each cu type ☼, ▲, ● are given in the rectangle).

$cu\_t_j \in \{cu\_t_j\}$ in the proper discrete time ($dtu$).

The value $|\{cu\_t_j\}|$ is equal to the number of computational units of *j*-type available during design process, e.g. for the graph in Fig.1 $\{cu\_t_j\}=\{☼,☼\}$. Each edge $e_{ij} \in E$ represents a data set. The process of assigning $v_i \in V$ onto $cu\_t_j$ is constrained by a limited set of resources ($\{cu\_t_j\} \cup ... \cup \{cu\_t_j\} \cup ...$) and the maximal number of time slots $dtu_M$.

Each type of processing unit is capable of processing a specified type of computational tasks. The problem of assigning tasks to processors as specified here is NP complete [3].

While assigning $ct_i$ to $cu\_t_j$ in BPD algorithm, different

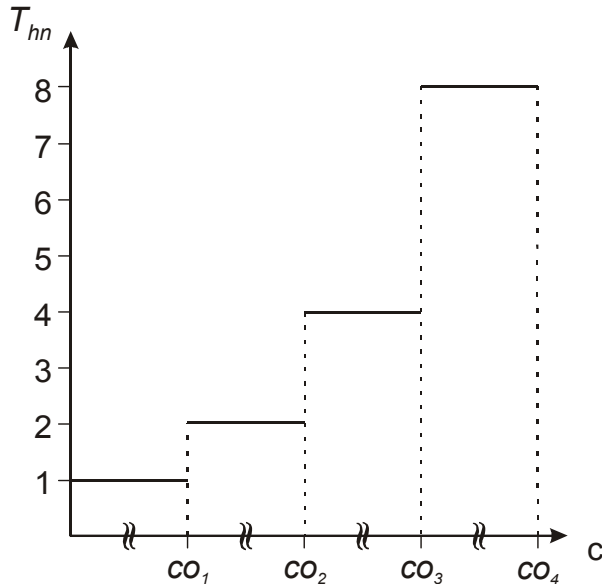

Fig. 2. Normalized throughput ($T_{hn}$) of different computational units versus their cost ($co_i$).

The second stage of BPD algorithm (Tab.I) consists of delaying the initial $SG$ created in the first stage. The $C_{su}$ set includes all the rows suitable for delaying, and $C_{ms}$ indicates the most suitable row selected from this set. The chosen $C_{ms}$ row actually undergoes the process of delaying.

TABLE I
PSEUDOCODE OF THE SECOND STAGE OF BPD ALGORITHM

```
1. C_su = SG
2. while ( |C_su| > 0 )
3.    foreach ( C_k ∈ C_su )
4.       if ( ! fsc_fulfiled( C_k ) )
5.          C_su = C_su \ C_k
6.    if ( |C_su| == 0 )
7.       break
8.    if ( |C_su| > 1 )
9.       foreach ( C_k ∈ C_su )
10.         if ( there_is_same_cu( C_k ))
11.            C_i = row_of_the_same_cu(C_k)
12.            foreach ( v_i ∈ C_k )
13.               if (fvi(v_j)>fvi( v_i ) )
14.                  interchange( v_i, v_j )
15.         C_ms = argmax_{C_k ∈ C_su} fck( C_k )
16.    else
17.       C_ms = HEAD( C_su )
18.    SG_backup = SG
19.    foreach ( v_i ∈ C_ms )
20.       insert_idle_tasks(v_i)
21.       if( ! delay_all_successors(v_i) )
22.          SG = SG_backup
23.          C_su = C_su \ C_ms
24.          break
```

values of throughput ($T_h$) of each $cu\_t_j$ are taken into consideration. The function describing normalized throughput ($T_{hn_i} = T_{h_i} / min_j T_{h_j}$) of different computational units versus their costs is shown in Fig. 2, where $0 < co_1 < co_2 < co_3 < co_4$. The appropriate cost of each computational unit characterized by the given throughput varies from $co_{Low}$ to $co_{High}$. The number of different levels of $T_{hn}$ results from the set of the computational units available for the designed system.

The main aim of the elaborated BPD algorithm is fulfilling the condition of providing the balanced power dissipation of computational units ($\{cu\_t_j\} \cup ... \cup \{cu\_t_j\} \cup ...)_T$) leading to reduction of average and peak temperatures of the digital VLSI CMOS system without deteriorating its throughput. In this process the total cost of the system is also decreased. The assumption is that less effective computational units are cheaper, so that replacing the chosen $cu\_t_i$ with other, less effective, results in decreasing the total system cost.

## IV. ALGORITHM DESCRIPTION

The elaborated BPD heuristic algorithm is partially based on the research results presented in papers [9] and [10] and concerned with reducing the power consumption of digital CMOS circuits. The algorithm consists of two stages described below. At the first stage of the algorithm the computational tasks represented by graph $G_T$ have to be assigned to the elements of resources with the lowest number of discrete time units taken into account. During this stage, either ASAP or ALAP [8] algorithm is executed.

Scheduled computational task for the example in Fig.1 after the first stage of BPD algorithm is shown in Fig.3.

**fsc_fulfiled**( $C_k$ ) (line 4)

This function performs the check for the *free space condition (fsc)*, defined by the following formula:

$$n_i \cdot l_k \le r_o \qquad (1)$$

where $n_i$ is the number of *dtu* needed to perform the task, $l_k$ is the number of tasks assigned to $C_k$ computational unit row, $r_o$ is the number of free task slots after the first occurrence of a task in $C_k$ computational unit row.

Formula (1) checks if the number of free *dtu* slots is sufficient for the idle tasks to introduce longer processing time of a cui. Every $C_k$ selected for the increased number of *dtu* has to fulfil condition (1). Despite the fact, that cascading tasks from the other $C_k$ 's are not taken into consideration while calculating *fsc*, condition (1) is sufficient to pre-reject quickly some $C_k$ from the $C_{su}$ set before starting the time-consuming delaying process.

**there_is_same_cu**($C_k$) (line 10)

This function simply indicates whether there is another $cu_i$ of exactly the same type as the one assigned to $C_k$, i.e. being capable of performing the same type of task in the same time.

**row_of_the_same_cu**($C_k$) (line 11)

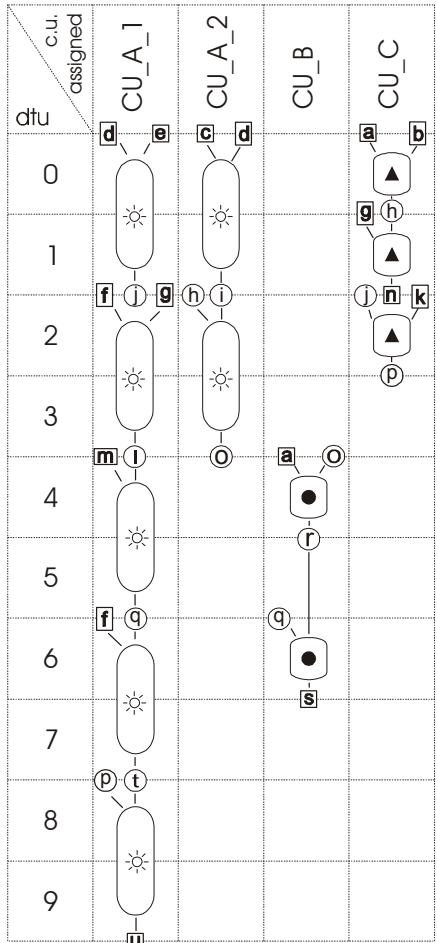

Fig. 3. Results (represented as the scheduling graph *SG*) of assignment computational task represented by graph $G_T$ in Fig.1 for resources {☼,☼, ▲, ●} obtained after the first stage of BPD algorithm.

This function returns a row containing the tasks of exactly the same type as $C_k$.

**fvi(** $v_i$ **)** (line 16)

This function calculates *fvi* factor for the $v_i$ task according to the following formula:

$$f_{vi} = \frac{i_{vi} + o_{vi} + (f_{avi} - s_{vi} \cdot n_i)}{p_i + 1} \qquad (2)$$

where $i_{vi}$ is the number of independent inputs of task $v_i$, $o_{vi}$ is the number of system outputs of task $v_i$,

$f_{avi} = dtu_M - (cs_e + n_i)$, $dtu_M$ is the maximal number of *dtu* admissible, $cs_e$ is the number of *dtu* which $v_i$ is assigned to, $s_{vi}$ is the number of tasks of the same type as task $v_i$ in the path of $G_T$ below task $v_i$, $n_i$ is the number of *dtu* needed to perform the task, $p_i$ is the *p* - label of task $v_i$, the minimal *p* - label of a task equals *0*, hence addition of *1* in the denominator is necessary to avoid dividing by *0*.

The $f_{vi}$ value of a task indicates its suitability for being slowed down. When there is more than one row of the same type it is used to create the best interconnect rows by interchanging tasks.

**interchange(** $v_i$, $v_j$ **)** (line 14)

This function swaps the $v_i$ and $v_j$ tasks, so that $v_i$ is located in task slots earlier occupied by $v_j$ and vice versa.

**fck(** $C_k$ **)** (line 15)

The value of the function is given by:

$$f_{Ck} = P_{dC_k}^n \cdot l_{Ck} \qquad (3)$$

where $P_{dCk}^n$ is the normalised computational load of $C_k$ computational unit row (cu-assigned to $C_k$ row, $P_{dCk}^n$ is normalised to $cu_i$, having the lowest value of $P_{di}$), $l_{Ck}$ is a number of tasks in $C_k$ computational unit row.

The *fck* function is responsible for selecting the most suitable row ($C_{ms}$) for inserting idle tasks, from the $C_{su}$ set. It chooses the row assigned to $cu\_ti$ that has the highest throughput demand, hence it gives the highest throughput demand reduction when the processing element $p_y^x$ assigned to row $C_k$ is slowed down.

**insert_idle_tasks(** $v_i$ **)** (line 20)

This function simply adds new task slots with idle tasks after the $v_i$ task. If there is an empty task slot after the last *dtu* occupied by $v_i$, then an idle task is added there. However, when there is no empty room for a new idle task, then the next task in the row of $v_i$ is delayed. Next the data interconnections between $v_i$ and its successor tasks must be checked. This is done by the *delay_all_successors* function described below.

**delay_all_successors(** $v_i$ **)** (line 21)

This function checks if all the data needed to perform successor task ($s_i$) of $v_i$ is available on time, by checking the condition:

$$end\_dtu(v_i) \leq start\_step(s_i) \qquad (4)$$

If it is not fulfilled, then the successor is delayed as many *dtu* as needed, so that *start_step($s_i$)=end_dtu($v_i$)*.

Such a delay implies the need for checking all the sets of data and interconnections between the successors of $s_i$. If the delay is not possible due to the $dtu_M$ constraint, increasing the number of *dtu* of $v_i$ (and the computational unit it is assigned to) fails.

In such a case the row containing $v_i$, i.e. $C_{ms}$ is removed from the $C_{su}$ set *F*, and the process starts from the beginning with the decreased $|C_{su}|$.

For a simple benchmark shown in Fig. 1, the results are presented in details. The obtained results are shown in Fig. 3 in a form of scheduling graphs *SG* for the first stage of the directly proportional to its throughput. To simplify the comparison of computational efficiency we assume that $T_{hi}$ can be lowered by the factor *0.5*. Table II presents the assumed normalized cost due to the throughput of each computational unit type.

TABLE II
THE ASSUMED NORMALIZED COST DUE TO THE THROUGHPUT OF EACH COMPUTATIONAL UNIT TYPE

| | cu_a | | | | cu_b | | | | cu_c | | | | cu_d | | | | cu_e | | | |
|---|---|---|---|---|---|---|---|---|---|---|---|---|---|---|---|---|---|---|---|---|
| $T_{hi}$ | 8 | 4 | 2 | 1 | 8 | 4 | 2 | 1 | 8 | 4 | 2 | 1 | 8 | 4 | 2 | 1 | 8 | 4 | 2 | 1 |
| $co$ | 8 | 4 | 2 | 1 | 8 | 4 | 2 | 1 | 8 | 4 | 2 | 1 | 8 | 4 | 2 | 1 | 8 | 4 | 2 | 1 |

TABLE III
THE ASSUMED NORMALIZED POWER DISSIPATION DUE TO THE THROUGHPUT OF EACH COMPUTATIONAL UNIT TYPE

| | cu_a | | | | cu_b | | | | cu_c | | | | cu_d | | | | cu_e | | | |
|---|---|---|---|---|---|---|---|---|---|---|---|---|---|---|---|---|---|---|---|---|
| $T_{hi}$ | 8 | 4 | 2 | 1 | 8 | 4 | 2 | 1 | 8 | 4 | 2 | 1 | 8 | 4 | 2 | 1 | 8 | 4 | 2 | 1 |
| $P_{di}$ | 8 | 4 | 2 | 1 | 8 | 4 | 2 | 1 | 8 | 4 | 2 | 1 | 8 | 4 | 2 | 1 | 8 | 4 | 2 | 1 |

Tables III and IV and Fig. 5 show number of computational units of each type before and after applying BPD algorithm, respectively.

TABLE IV
NUMBER OF COMPUTATIONAL UNITS OF EACH TYPE BEFORE APPLYING BPD ALGORITHM

| | cu_a | | | | cu_b | | | | cu_c | | | |
|---|---|---|---|---|---|---|---|---|---|---|---|---|
| $T_{hi}$ | 8 | 4 | 2 | 1 | 8 | 4 | 2 | 1 | 8 | 4 | 2 | 1 |
| s298 | 17 | | | | 5 | | | | 10 | | | |
| c5315 | 158 | | | | 56 | | | | 68 | | | |
| s382 | 4 | | | | 9 | | | | 7 | | | |
| s444 | 4 | | | | 17 | | | | 7 | | | |
| s526 | 32 | | | | 9 | | | | 14 | | | |
| s5378 | | | | | | | | | 47 | | | |

| | cu_d | | | | cu_e | | | |
|---|---|---|---|---|---|---|---|---|
| $T_{hi}$ | 8 | 4 | 2 | 1 | 8 | 4 | 2 | 1 |
| s298 | 7 | | | | 31 | | | |
| c5315 | 7 | | | | 169 | | | |
| s382 | 6 | | | | 30 | | | |
| s444 | 8 | | | | 30 | | | |
| s526 | 14 | | | | 38 | | | |
| s5378 | 119 | | | | 333 | | | |

TABLE V
NUMBER OF COMPUTATIONAL UNITS OF EACH TYPE AFTER APPLYING BPD ALGORITHM

| | cu_a | | | | cu_b | | | | cu_c | | | |
|---|---|---|---|---|---|---|---|---|---|---|---|---|
| $T_{hi}$ | 8 | 4 | 2 | 1 | 8 | 4 | 2 | 1 | 8 | 4 | 2 | 1 |
| s298 | 3 | 11 | 3 | | 2 | 3 | | | 3 | 2 | 5 | |
| c5315 | 100 | 12 | 8 | | 38 | 53 | | | 3 | 62 | 2 | 4 |
| s382 | 3 | 1 | | | 7 | 1 | 1 | | 7 | | | |
| s444 | 2 | 2 | | | 11 | 6 | | | 2 | 1 | 1 | 3 |
| s526 | 6 | 18 | 8 | | 3 | 4 | 2 | | 1 | 10 | 3 | |
| s5378 | | | | | | | | | 28 | 18 | 1 | |

| | cu_d | | | | cu_e | | | |
|---|---|---|---|---|---|---|---|---|
| $T_{hi}$ | 8 | 4 | 2 | 1 | 8 | 4 | 2 | 1 |
| s298 | 3 | 3 | 1 | | 23 | 2 | 5 | 1 |
| c5315 | 7 | | | | 130 | 14 | 3 | 22 |
| s382 | 5 | 1 | | | 13 | 9 | 5 | 3 |
| s444 | 8 | | | | 18 | | 7 | 5 |
| s526 | 7 | 6 | 1 | | 9 | 22 | 6 | 1 |
| s5378 | 93 | 1 | 4 | 21 | 285 | 20 | 4 | 24 |

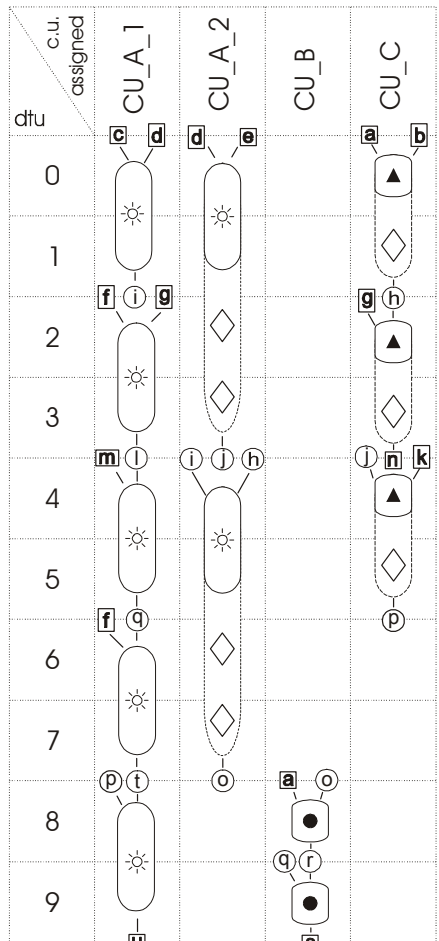

Fig. 4. Results (represented as the scheduling graph SG) of assignment computational task represented by graph GT in Fig.1 for resources {☼, ☼, ▲, ●} obtained after the second stage of BPD algorithm (the symbol ◊ describes decreasing the computational efficiency of the chosen computational unit).

algorithm, while the second stage is given in Fig. 4.

There in Fig. 4 lowering the cost of the appropriate computational units is represented by inserting the symbol ◊. It means that its throughput can be twice as low without deteriorating the efficiency of the whole computational system. Therefore our example for computational units *cu_a_2* and *cu_c* show that their throughput can be lowered twice resulting in cost reduction of the designed computational system. Moreover, the value of the throughput obtained earlier does not deteriorate.

## IV.  EXPERIMENTAL RESULTS

This section presents the results obtained by applying the BPD algorithm on selected benchmarks [12].

Cost reduction is calculated for each computational task based on the number of computational units of each type before and after application of BPD algorithm. The cost of each computational unit type assumed for cost calculation is

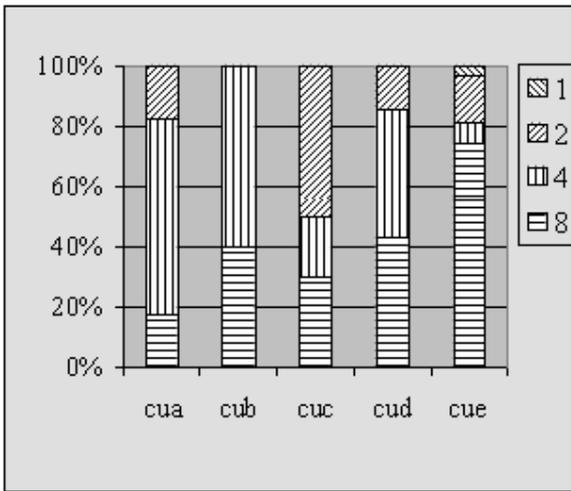

(a)

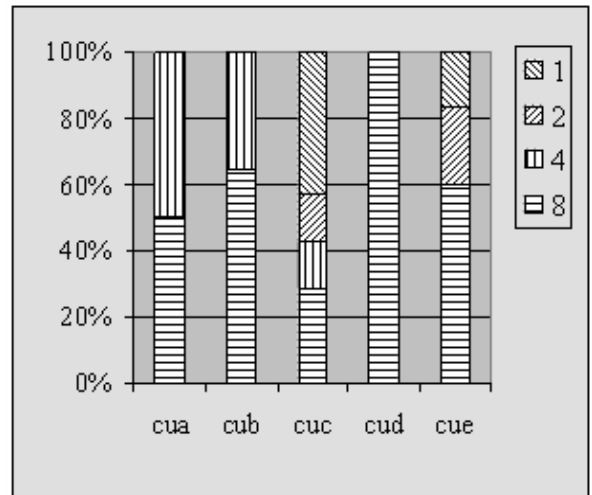

(d)

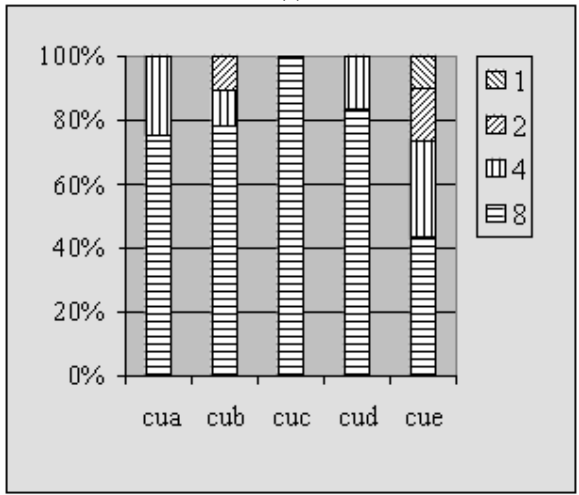

(b)

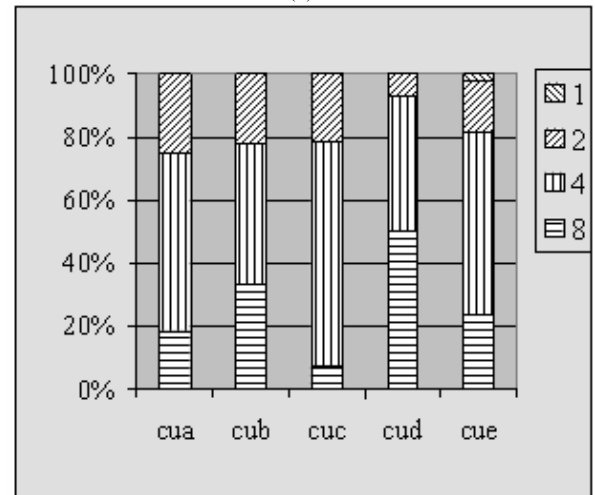

(e)

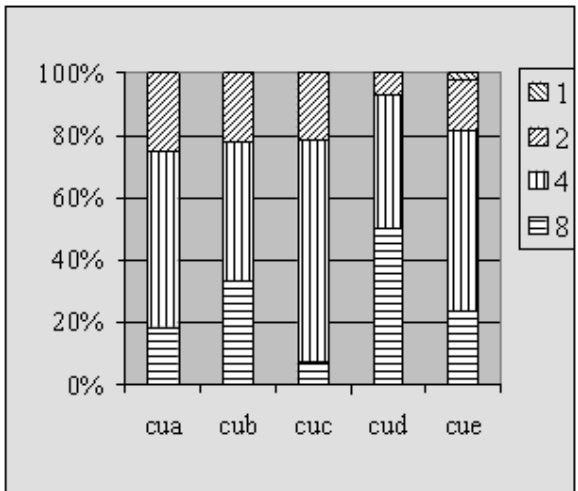

(c)

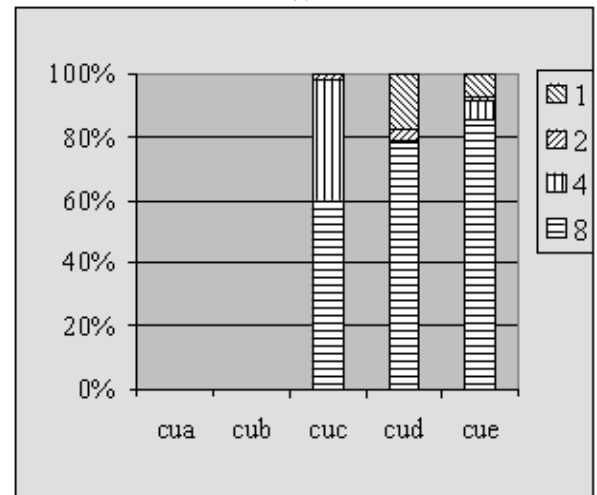

(f)

Fig. 5. Percentage of computational units of each type after applying BPD algorithm for s298 (a), c5315 (b), s382 (c), s444 (d), s526 (e), s5378 (f).

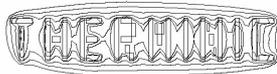

TABLE VI
EXPERIMENTAL RESULTS

| Benchmark name | Number of graph vertices | Cost reduction % |
|---|---|---|
| s298 | 119 | 31.25 |
| c5315 | 1994 | 17.66 |
| s382 | 158 | 23.44 |
| s444 | 181 | 26.52 |
| s526 | 193 | 42.87 |
| s5378 | 2779 | 13.15 |

The resulting cost reduction together with the number of the $G_T$ graph vertices of each benchmark computational task is reported in Table VI.

## V. CONCLUSIONS

In this paper the BPD heuristic scheduling algorithm for load balanced power dissipation resulting in reduction of average and peak temperatures of the digital VLSI CMOS digital circuits/systems was presented. The objective function introduced is measured by cost reduction of VLSI CMOS digital circuits which directly depends on power dissipated in IC. The main idea of BPD algorithm is based on decreasing the cost of chosen computational units by adjusting their efficiency to real needs without deterioration of the computational efficiency of the whole system.

The applied BPD algorithm has been verified for the chosen set of benchmarks. Experimental results proved 13 to 43 per cent cost reduction of the computing system achieved without deterioration of the system throughput with the assumed cost to throughput dependency. This reduction has a straightforward influence on decreasing the average and peak temperatures of VLSI CMOS system and results in increasing its reliability.

## ACKNOWLEDGMENT

The work was partially supported by the State Committee for Scientific Research (MNiI, Poland) through the grant 3 T11B 015 27.